# Comment on "Erratum: "Two-dimensional porous graphitic carbon nitride $C_6N_7$ monolayer: First-principles calculations" [Appl. Phys. Lett. 119, 142102 (2021)]"

Bohayra Mortazavi[a,*], Fazel Shojaei[b] and and Masoud Shahrokhi[c]

[a]Department of Mathematics and Physics, Leibniz Universität Hannover, 30167 Hannover, Germany.
[b]Faculty of Nano and Bioscience and Technology, Persian Gulf University, Bushehr 75169, Iran.
[c]Independent Researcher, Lyin, France.

*Corresponding author: bohayra.mortazavi@gmail.com;

Recently, Bafekry et al. [*Appl. Phys. Lett. 120, 189901* (**2022**)] reported their density functional theory (DFT) results on the elastic constants of $C_6N_7$ monolayer. They predicted non-zero elastic constants along the out-of-plane direction for a single-layered material, which contradicts with basic physics of the stiffness tensor for plane stress condition. Moreover, in their work Young's modulus is erroneously calculated. On the basis of DFT calculations, herein we predicted the $C_{11}$, $C_{12}$ and $C_{66}$ of the $C_6N_7$ monolayer to be 286, 73 and 107 GPa, respectively, equivalent with an in-plane Young's modulus of 267 GPa. Using DFT calculations and a machine learning interatomic potential, we also show that $C_6N_7$ monolayer shows isotropic elasticity.

Bafekry et al.[1] predicted the $C_{11}$ = 258.6 GPa, $C_{22}$ = 290.8 GPa, and $C_{12}$ = 70.73 GPa and $C_{13}$ = $C_{23}$ = 2.49 GPa, $C_{33}$ = 9.05 GPa, $C_{44}$ = 25.86 GPa, $C_{55}$ = 2.90 GPa, and $C_{66}$ = 3.08 GPa for the $C_6N_7$ monolayer. An isolated layer that is in contact with vacuum on the both sides and is thus free to move toward the out-of-plane direction upon the geometry optimization, follow the plane-stress theory and cannot exhibit non-zero $C_{13}$, $C_{23}$, $C_{33}$, $C_{44}$ and $C_{55}$ values. In contrast with basic physics, Bafekry and coworkers in their original work[2] and their recently published erratum.[1], reported $C_{13}$, $C_{23}$, $C_{33}$, $C_{44}$ and $C_{55}$ values for the $C_6N_7$ monolayer with two-decimal precisions. In contrast with data in their original work[2] and our previous comment[3], Ref. [1] reported different $C_{11}$ and $C_{22}$ values for the for the $C_6N_7$ monolayer, which suggests that this material shows an anisotropic elasticity. For an anisotropic material, Young's modulus and other transport properties become direction-dependent, however Bafekry et al.[1] erroneously reported a single Young's modulus for this system. On the basis of Hooke's law, the Young's modulus of a 2D anisotropic material follows the following relations with respect to the elastic constants:

$$Y_{zig} = \frac{C_{11} \times C_{22} - C_{12}^2}{C_{22}} \quad (1)$$

$$Y_{arm} = \frac{C_{11} \times C_{22} - C_{12}^2}{C_{11}} \quad (2)$$

Here, $Y_{zig}$ and $Y_{arm}$ are the Young's modulus along the zigzag (direction 1) and armchair (direction 2) directions (find Fig. 1a inset), respectively. According to the aforementioned relations and on the basis of Bafekry *et al.*[1] data, $Y_{zig}$ and $Y_{arm}$ are obtained to be 241.4 and 271.5 GPa, respectively. They however erroneously reported a Young's modulus of 362.9 GPa for the $C_6N_7$ monolayer, which is not consistent with their own calculated elastic constants. Basically the Young's modulus cannot take a value higher than both $C_{11}$ and $C_{22}$ values.

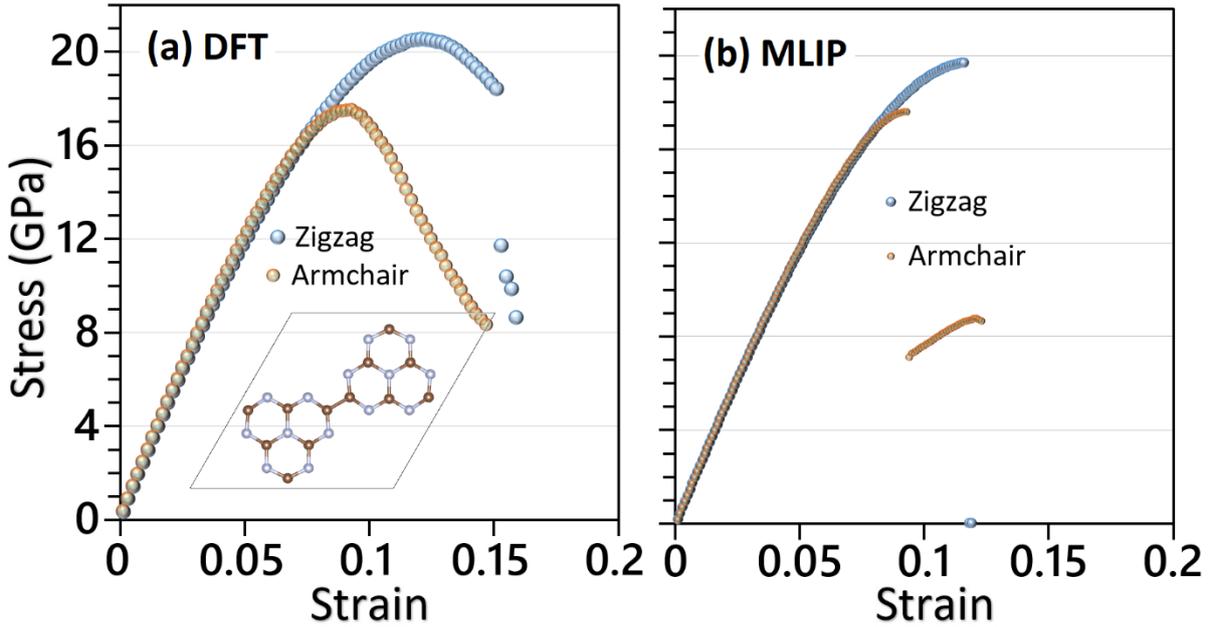

**Fig. 1**, Uniaxial stress-strain responses of the $C_6N_7$ monolayer along the armchair and zigzag directions predicted by (a) DFT and (b) a machine learning interatomic potential (MLIP). Inset in panel (a) shows the crystal structure of the $C_6N_7$ monolayer, where horizontal and vertical direction correspond to the zigzag and armchair directions, respectively.

Using the DFT calculations and with assuming a thickness of 3.35 Å for the $C_6N_7$ monolayer (using that of the graphene), we predicted the $C_{11}$, $C_{12}$ and $C_{66}$ of the $C_6N_7$ monolayer to be 286, 73 and 107 GPa, respectively, equivalent with an in-plane Young's modulus of 267 GPa. Our predicted $C_{66}$ for the $C_6N_7$ monolayer is by almost 35 folds higher than that reported in Ref.[1]. To provide a complete vision on the mechanical response of the $C_6N_7$ monolayer, we recalculated the uniaxial stress-strain responses along the armchair and zigzag directions using the DFT and a machine learning interatomic potential (MLIP). The computational details for these simulations are the same as those in our recent study [4], however finer strain steps are used to enhance the resolution. As it is clear, with both methods the stress-strain curves along the armchair and zigzag directions coincide for the initial linear section, confirming isotropic elasticity in this system.

In conclusion, the elastic constants and Young's modulus of the $C_6N_7$ monolayer predicted by Bafekry and coworkers in their original work[2] and their recently published erratum[1] are erroneous and contradict with Hooke's law for the plane stress condition .

## Acknowledgments

B.M. appreciates the funding by the Deutsche Forschungsgemeinschaft (DFG, German Research Foundation) under Germany's Excellence Strategy within the Cluster of Excellence PhoenixD (EXC 2122, Project ID 390833453). F.S. thanks the Persian Gulf University Research Council, Iran for support of this study. B. M is greatly thankful to the VEGAS cluster at Bauhaus University of Weimar for providing the computational resources.